\documentclass[useAMS,usenatbib]{mn2e}
\usepackage{epsf, epsfig}

\title[New flaring of an ultraluminous X-ray source in NGC\,1365]
{New flaring of an ultraluminous X-ray source in NGC\,1365}
\author[R. Soria et al.]
{R. Soria$^{1,2}$\thanks{E-mail: rsoria@cfa.harvard.edu}, 
A. Baldi$^{1}$,
G. Risaliti$^{1,3}$,
G. Fabbiano$^{1}$,
A. King$^{4}$,
V. La Parola$^{5}$,
A. Zezas$^{1}$
\\
$^{1}$\/Harvard-Smithsonian Center for Astrophysics, 
	60 Garden st, Cambridge, MA 02138, USA\\
$^{2}$\/Mullard Space Science Laboratory (UCL), Holmbury St Mary, 
	Dorking, Surrey, RH5 6NT, UK\\
$^{3}$\/INAF, Osservatorio di Arcetri, Lgo Fermi 5, 50125 Firenze, Italy\\
$^{4}$\/Department of Physics and Astronomy, University of Leicester, 
	LE1 7RH, UK\\
$^{5}$\/INAF, Istituto di Astrofisica Spaziale e Fisica Cosmica, Palermo, Italy}

\begin{document}

\date{Received 05 Feb 2007}

\pagerange{\pageref{firstpage}--\pageref{lastpage}} \pubyear{2006}

\maketitle

\label{firstpage}

\begin{abstract}
We have studied a highly variable ultraluminous X-ray source (ULX) 
in the Fornax galaxy NGC\,1365, with a series of 12 
{\it Chandra} and {\it XMM-Newton} 
observations between 2002 and 2006. In 2006 April, the source peaked 
at a luminosity $\approx 3 \times 10^{40}$ erg s$^{-1}$ 
in the $0.3$--$10$ keV band (similar to the maximum luminosity 
found by {\it ASCA} in 1995), and declined  
on an $e$-folding timescale $\approx 3$ days. The X-ray spectrum 
is always dominated by a broad power-law-like component. 
When the source is seen at X-ray luminosities $\approx 10^{40}$ erg s$^{-1}$, 
an additional soft thermal component (which we interpret 
as emission from the accretion disk) contributes $\approx 1/4$ 
of the X-ray flux; when the luminosity is higher, 
$\approx 3 \times 10^{40}$ erg s$^{-1}$, the thermal component 
is not detected and must contribute $< 10\%$ of the flux.
At the beginning of the decline, ionized absorption 
is detected around $\sim 0.5$--$2$ keV; it is a possible 
signature of a massive outflow.
The power-law is always hard, with a photon index $\Gamma \approx 1.7$ 
(and even flatter at times), as is generally the case with bright 
ULXs. We speculate that this source and perhaps most other bright ULXs 
are in a high/hard state: as the accretion rate increases 
well above the Eddington limit, more and more power is extracted 
from the inner region of the inflow through non-radiative 
channels, and is used to power a Comptonizing corona, jet or wind. 
The observed thermal component comes from the standard outer disk; 
the transition radius between outer standard disk and Comptonizing 
inner region moves further out and to lower disk temperatures 
as the accretion rate increases. This produces 
the observed appearance of a large, cool disk. 
Based on X-ray luminosity and spectral arguments, 
we suggest that this accreting black hole has 
a likely mass $\sim 50$--$150 M_{\odot}$ (even without 
accounting for possible beaming).

\end{abstract}


\section{Introduction: ULX variability}

It is still not clear how ultraluminous X-ray sources 
(ULXs) relate to the class of stellar-mass black holes (BHs) 
first identified and extensively studied in the Galaxy 
and Large Magellanic Cloud; in particular, whether 
and how they differ from high-mass X-ray binaries 
(systems where the donor is an OB star).  They could  
represent the upper end of the BH mass distribution, 
perhaps up to $\sim 100 M_{\odot}$. Alternatively, 
the emission could be mildly beamed in our direction, 
enhancing the apparent luminosity perhaps by a factor 
of a few. Finally, the emission could be super-Eddington 
by a factor of a few. One of the reasons the ULX 
classification remains controversial is that very little 
is known about their spectral state transitions.
The identification and interpretation 
of state transitions over characteristic timescales 
of a few months/years has provided a powerful tool 
to understand the accretion processes (for 
example the disk/corona/jet interplay as a function 
of a variable mass accretion rate) in Galactic BHs 
over the last two decades (Remillard \& McClintock 2006). 

Because of their distance, 
ULXs are too faint to be regularly observed by the current 
generation of X-ray all-sky monitors, and require 
targeted observations. Few ULXs have been observed  
regularly enough to determine their long-term 
X-ray behaviour; most of them may vary by a factor 
of a few (compare for example the catalogues 
of Swartz et al.~2004 and Liu \& Bregman 2005) 
but tend to be persistently in a bright state.
From this point of view, they are more 
similar to the (few) Galactic BHs in high-mass 
binary systems (i.e., accreting from an OB donor star) 
than to the more numerous ones in low-mass systems 
(accreting from a low-mass star).
Significant variability or on/off transitions 
have been monitored in some of the fainter ULXs
($L_{\rm x} \approx$ a few $10^{39}$ erg s$^{-1}$), 
for example in the Antennae (Fabbiano et al. 2003).
However, spectral state transition or flux variations 
by more than one order of magnitude are rarer among 
the brightest systems ($L_{\rm x} > 10^{40}$ erg s$^{-1}$).
Among the few notable exceptions (apart from NGC\,1365 X1, 
object of this paper) are the two brightest ULXs in M\,82 
(Kaaret, Simet \& Lang 2006), the brightest ULX 
in the Cartwheel (Wolter, Trinchieri \& Colpi~2006), a source  
in NGC\,3628 (Strickland et al. 2001), 
and a peculiar super-soft source that underwent an outburst 
in M\,101 (Kong \& Di Stefano 2005). The last source may or 
(more likely) may not belong to the same physical class 
as most other ULXs.

In the few ULXs where the flux and/or spectral variations are 
significant enough to be considered a state transition, 
the pattern does not appear to correspond to the ``canonical'' 
sequence of physical states displayed by stellar-mass BHs: 
quiescence; low/hard state (hard power-law);
thermal-dominant state (disk-blackbody with 
$kT_{\rm in} \sim 1$ keV); very high state state (softer 
power-law plus disk-blackbody component)
(Remillard \& McClintock 2006; Fender, Belloni \& Gallo~2004).
Instead, ULX spectra in the $0.3$--$10$ keV band seem to show 
three phenomenological features 
(Stobbart, Roberts \& Wilms~2006), which may or may not 
correspond to truly different physical components: a) a broad, 
power-law-like component with photon index 
$\Gamma \approx 1.7$--$2.3$; b) a curvature or break 
above $\sim 5$ keV; c) an additional 
soft component ($kT_{\rm in} \approx 0.15$--$0.25$ keV).
Strong flux changes generally correspond to one or two 
of those features becoming more prominent or disappearing.
For example, at higher luminosities, the power-law component 
appears to increase more than the soft thermal 
component; the high-energy steepening 
is also more noticeable at higher luminosities
(e.g., Roberts 2007; Dewangan et al.~2004; Kubota et al.~2001).
It is difficult to classify this behaviour in terms 
of state transitions, because there is still no consensus 
on the physical origin of those three basic features.
In the standard scenario, the soft component 
is emitted by a cool disk, the power-law component 
is the result of Comptonization, and its curvature/break is due 
to an increased optical depth of the Comptonizing region.
Alternatively, it was suggested that the curved emission 
component may come from a hot accretion disk (e.g., 
Kubota et al.~2001; Roberts et al.~2005; Stobbart et al.~2006; 
Tsunoda et al.~2006).

For our study, we have targeted the brightest ULX 
in NGC\,1365 (henceforth, X1, following the {\it ROSAT} 
identification; Komossa \& Schulz 1998) 
which was already known to be highly 
variable (a factor $\ga 10$) over a few months' timescale, 
from {\it ROSAT} and {\it ASCA} studies 
(Komossa \& Schulz 1998; Iyomoto et al.~1997).
The source is located in the most luminous spiral galaxy 
of the Fornax Cluster, a supergiant barred spiral 
(type SBb(s)I: Sandage \& Tammann 1981) with 
a total gas mass $\approx 3 \times 10^{10} M_{\odot}$, 
a kinematic mass $\approx 3.6 \times 10^{11} M_{\odot}$, 
and a star formation rate $\sim 10 M_{\odot}$ yr$^{-1}$ 
(Roussel et al.~2001; Lindblad 1999).
The Cepheid distance to NGC\,1365 is $19 \pm 1$ Mpc 
(Ferrarese et al.~2000); at that distance, $1\arcsec = 92$ pc.   
In this paper, we will show that this source varies 
dramatically in luminosity over short timescales, and 
that such changes are associated to spectral changes, 
including the signature of an ionized outflow. 
We will then discuss whether and how those changes 
may be consistent with a standard disk plus corona 
scenario, and what constraints we can obtain 
on the BH mass and accretion rate.

\section{Summary of our observations}

We observed the central region of NGC\,1365 with 
{\it Chandra} on 2002 December 24, then with {\it XMM-Newton}
on five occasions between 2003 January 16 and 2004 July 24, 
and with {\it Chandra} again six times between 
2006 April 10 and 2006 April 23 (see Table 1 for details).
Henceforth, for simplicity, we label our {\it XMM-Newton}/EPIC 
and {\it Chandra}/ACIS observations as E1--E5, 
and A1--A6, respectively, in chronological order.
The main target was the active nucleus: findings 
for that study are reported elsewhere (Risaliti et al.~2005a,b; 
Risaliti~2007; Risaliti et al.~2007 in preparation).

\subsection{{\it XMM-Newton} observations}

For all the {\it XMM-Newton} observations, the European 
Photon Imaging Camera (EPIC) was in full-window mode, 
with the medium filter, both for the pn and the 
two Metal Oxide Semi-conductor (MOS) detectors. 
We used the Science Analysis System 
({\small SAS}) version 5.4.3 (release date 2003-09-11) 
for the EPIC spectral extraction and analysis, 
but we also checked that the results were unchanged 
when using the more recent {\small SAS} version 7.0.0 
(2006-06-28). We selected only unflagged events 
(\#XMMEA\_EP for the pn and \#XMMEA\_EM for the
two MOSs), with pattern $\leq$ 12 for the MOSs, 
pattern $\leq$ 4 for the pn.
We defined good-time-interval files, removing intervals 
of high background. We extracted spectra and lightcurves 
for X1 using the {\small SAS} task {\it xmmselect}: 
the source extraction region was a circle of radius 
$20\arcsec$, which includes $\approx 72\%$ of the energy 
in the pn and $\approx 75\%$ in the MOSs.
The background was extracted from a circular region 
of radius $60\arcsec$, located at the same angular 
distance from the nucleus of the galaxy.
We built response and auxiliary response files 
with the {\small SAS} tasks {\it rmfgen} and 
{\it arfgen} respectively. For each EPIC observation, 
we then coadded the pn and MOS spectra, with the method 
described in Page, Davis \& Salvi (2003), 
in order to increase the signal-to-noise ratio.
Finally, we fitted the background-subtracted spectra 
with standard models in {\small XSPEC} version 11.3.1 
(Arnaud 1996). For the X-ray timing analysis, we 
used the {\small SAS} version 7.0.0. We used {\it lccorr} 
to correct the background-subtracted lightcurves 
for gaps in the good-time-interval. We then used standard 
{\small XRONOS} tasks for studying the lightcurves.

In addition, Optical Monitor (OM) images were taken in 
the $U$, $UVW1$, $UVM2$ and $UVW2$ filters\footnote{See 
http://www.xmm.ac.uk/onlines/uhb/XMM\_UHB/node66.html 
for a plot of the OM filter throughput curves.}, 
during most of the observations (Table 1). 
We registered the astrometry of each 
OM image to a sample of bright stars in 
the USNO-B1.0 Catalog (Monet et al.~2003), 
so that the positional error is $\la 0\farcs6$ 
for each image, comparable to the uncertainty 
in the {\it Chandra} astrometry. 
We used standard {\small IRAF} tasks 
to combine images in the same filter, 
from different observations. (In fact, 
the exposures in the $UVW2$ filters proved to be  
too short to provide useful information).

\subsection{{\it Chandra} observations}

{\it Chandra} observations were performed with the back-illuminated CCD
of the Advanced Camera for Imaging and Spectroscopy (ACIS-S, Weisskopf et 
al.~2002). The first observation was taken with the full frame; 
the others in the standard 1/4 subarray mode. The subarray mode 
enables a faster frame time (0.8~s instead of 3.2~s for full frame 
observations), thus reducing the pile up fraction. This is never 
a relevant issue for the analysis of the ULX; it was selected 
to prevent high pile up fractions in the main target, 
the active nucleus.

All {\it Chandra} datasets were analyzed with 
the Chandra Interactive Analysis of Observations 
({\small CIAO}) software, version 3.3.
For each observation, the source spectrum was extracted 
from a circular region of radius $2\arcsec$ centred on the
ULX position. The background spectrum was extracted from circular 
regions near the ULX. The response and ancillary response 
functions were created using standard {\small CIAO} tasks.
with the latest calibration files available.
The background-subtracted spectra were later re-binned 
to a signal-to-noise ratio $\ga 4$ and fitted
with standard models in {\small XSPEC} version 11.3.1 
(Arnaud 1996).

\section{Results of our X-ray study of X1}

We found three ULXs within a projected distance 
$\sim 15$ kpc from the active nucleus (Figures 1, 2, 3).
We unofficially designate them, for simplicity, X1, X2 and X3, 
in order of average X-ray brightness over the whole 
series of our observations. X1 is the ULX previously 
found by {\it ROSAT} and {\it ASCA}. A more accurate 
source position for X1, from the {\it Chandra} image, is 
RA = $03^h33^m34^s.61$, Dec $= -36^{\circ}09\arcmin36\farcs6$.
All three ULXs (but most remarkably, 
X1 and X3) vary significantly between different observations. 
In this paper, we discuss the main properties 
of X1, and leave a study of the other ULXs and 
non-nuclear X-ray sources in NGC\,1365 to a follow-up 
paper.

\subsection{Long- and short-term variability}

The long-term variability of X1 was already 
known from {\it ASCA} and {\it ROSAT} observations 
(Figure 4). As a first step, we fitted all 
the {\it Chandra} and {\it XMM-Newton} spectra
with absorbed power-law models (Table 2), 
confirming that its X-ray luminosity fluctuated 
by a factor of 10 over the few years of our 
observations. 
Our sequence of six {\it Chandra} observations 
between 2006 April 10--23 caught the rise, peak 
and decline of an outburst, in which the source 
reached unabsorbed X-ray luminosities of 
$(3.0 \pm 0.2) \times 10^{40}$ erg s$^{-1}$ 
in the $0.3$--$10$ keV band. This luminosity  
is similar to the peak observed by ASCA in 
1995 January, which we have re-estimated 
from the original {\it ASCA}/SIS data 
as $(3.0 \pm 0.5) \times 10^{40}$ erg s$^{-1}$ 
in the same band\footnote{There are inconsistencies 
between the {\it ASCA} count rates, fluxes 
and luminosities listed for X1 in different sections 
of Iyomoto et al.~1997. From the reported SIS 
count rate and fitted spectral parameters, 
we suggest that the unabsorbed luminosity 
must have been $\approx 3.2 \times 10^{40}$ 
erg s$^{-1}$ in the $0.3$--$10$ keV band, 
less than half of what is reported in Section 2.2.1 of
that paper. As a further check, we have re-extracted 
and re-analysed the {\it ASCA} data, confirming 
the lower value for the unabsorbed luminosity.}. 
It was considered remarkable 
at that time that the source must have risen and declined 
by more that an order of magnitude in less than a few weeks. 
In fact, the 2006 outburst shows that the characteristic 
timescale for the rise and decline is only a few days 
(Tables 1, 2). The peak of the outburst lasted for 
about a week, corresponding to the observations 
A3, A4 and A5 (2006 April 12--17).

Such long-term changes in the X-ray luminosity do not 
seem to correspond to significant variations 
in the slope of the continuum. When the spectra 
are fitted with an absorbed power-law model, the photon index 
remains $\sim 1.7$--$1.9$ over all our {\it XMM-Newton} 
and {\it Chandra} observations, consistent 
the value $\Gamma \approx 1.7$ that was estimated 
from the {\it ASCA} data during the 2005 January outburst 
(Iyomoto et al.~1997). Such slopes are similar 
to what is often found in other bright ULXs, but are 
flatter than the power-law slope in most Galactic BHs 
when they are in a high or very high state 
(Remillard \& McClintock 2006).
In fact, a more detailed spectral analysis based 
on two-component models (power-law plus disk-blackbody) 
reveals that the true slope of the power-law 
is in some cases even flatter, with a photon index $\approx 1$ 
(Section 3.2).

Only two of our observations (E4 and E5) are long enough 
to enable an analyis of short-term variability. 
We find that the count rate during E5 is consistent 
with being constant, while some flickering (quite common 
in accreting sources) is marginally
detected in E4 (Kolmogorov-Smirnoff probability of constancy 
$ = 1.5 \times 10^{-3}$). The total, soft-band and hard-band 
lightcurves for the EPIC-pn observation E4 are shown in Figure 4. 
We do not find any evidence of X-ray hardness changes 
during that observation, nor any significant features 
in its power density spectrum.

\begin{table*}
      \caption{Log of our observations of NGC\,1365. Henceforth, we
      shall use the lables A\# (ACIS) and E\# (EPIC) to identify the
      various observations, for simplicity.}
         \begin{tabular}{lcccccc}
            \hline
            \noalign{\smallskip}
Label & Date & MJD (start)
& Mission     
& Obs ID
                 & Good time interval (ks) & Optical Monitor\\
            \noalign{\smallskip}
            \hline
            \noalign{\smallskip}
A1 & 2002 Dec 24 & 52632.62 & {\it Chandra} ACIS-S & 3554 & 13.8  & \\[3pt]
E1 & 2003 Jan 16 & 52655.98 & {\it XMM-Newton} EPIC & 0151370101 &
            19.0 (MOS), 15.3 (pn) & $UVW2$ (3.5 ks)\\[3pt]
E2 & 2003 Feb 09 & 52679.86 & {\it XMM-Newton} EPIC & 0151370201 & 4.4
            (MOS), 2.5 (pn) & \\[3pt]
E3 & 2003 Aug 13 & 52864.01 & {\it XMM-Newton} EPIC & 0151370701 & 8.0
            (MOS), 6.4 (pn) & $UVW2$ (1.9 ks)\\[3pt]
E4 & 2004 Jan 17 & 53021.13 & {\it XMM-Newton} EPIC & 0205590301 &
            57.1 (MOS), 49.7 (pn) & $U$ (3.4 ks), $UVW1$ (3.4 ks), \\
	    &&&&&&$UVM2$ (3.4 ks)\\[3pt]
E5 & 2004 Jul 24 & 53210.27 & {\it XMM-Newton} EPIC & 0205590401 &
            64.3 (MOS), 55.6 (pn) & $U$ (3.8 ks), $UVW1$ (3.8 ks),\\
	    &&&&&&$UVM2$ (4.5 ks)\\[3pt]
A2 & 2006 Apr 10 & 53835.29 & {\it Chandra} ACIS-S & 6871 & 13.4  & \\[3pt]
A3 & 2006 Apr 12 & 53837.52 & {\it Chandra} ACIS-S & 6872 & 14.6  & \\[3pt]
A4 & 2006 Apr 15 & 53839.98 & {\it Chandra} ACIS-S & 6873 & 14.6  & \\[3pt]
A5 & 2006 Apr 17 & 53842.79 & {\it Chandra} ACIS-S & 6868 & 14.6  & \\[3pt]
A6 & 2006 Apr 20 & 53845.42 & {\it Chandra} ACIS-S & 6869 & 15.5  & \\[3pt]
A7 & 2006 Apr 23 & 53848.42 & {\it Chandra} ACIS-S & 6870 & 14.6  & \\
   \noalign{\smallskip}
            \hline
         \end{tabular}
         \label{t:allsources}
\end{table*}


\begin{table*}
      \caption{Instrumental count rates and emitted luminosities in
	the $0.3$--$10$ keV band, over
	our series of observations (the EPIC count rate is the sum 
	of pn and the two MOS's). Luminosities are based on the 
	best fitting power-law models, for which we also list 
	column densities, photon indices and normalizations 
	(in units of $10^{-5}$ photons keV$^{-1}$ cm$^{-2}$ 
	s$^{-1}$ at 1 keV). In addition, we assumed a Galactic 
	line-of-sight column density of $1.4 \times 10^{20}$ cm$^{-2}$ 
	for all fits. More complex fits were obtained for 
	some of the observations: see Tables 3--7.}
         \begin{tabular}{lcccccc}
            \hline
            \noalign{\smallskip}
Label & \multicolumn{2}{c}{Count rate} & \multicolumn{4}{c}{Power-law fit parameters}\\[3pt]
 & $10^{-2}$ ACIS-S cts s$^{-1}$ & $10^{-2}$ EPIC cts s$^{-1}$ 
& $N_{\rm H}$ ($10^{21}$ cm$^{-2}$) & Photon
index & $K_{\rm po} (10^{-5}) $ & $L_{0.3-10}$ ($10^{39}$ erg s$^{-1}$)\\  
            \noalign{\smallskip}
            \hline
            \noalign{\smallskip}
A1 &  $1.65 \pm 0.11$ & & $2.3^{+0.9}_{-1.0}$ & $2.12^{+0.49}_{-0.40}$ &
$3.2^{+1.5}_{-0.9}$ & $7.4^{+1.0}_{-0.6}$\\[5pt]
E1 &  & $2.61\pm 0.14$ & $0.2^{+0.5}_{-0.2}$ & $1.76^{+0.16}_{-0.24}$ &
$1.2^{+0.3}_{-0.2}$ & $3.3^{+0.4}_{-0.3}$ \\[5pt]
E2 &  & $1.73 \pm 0.25$ & $<1.8$ & $1.75^{+1.15}_{-0.37}$ & 
$0.8^{+0.4}_{-0.2}$ & $2.2^{+0.6}_{-0.2}$ \\[5pt]
E3 &  & $2.61\pm 0.22$ &  $0.6^{+1.3}_{-0.6}$ &
$2.27^{+0.48}_{-0.67}$ & $1.3^{+1.1}_{-0.4}$ & $2.9^{+0.5}_{-0.4}$\\[3pt]
E4 &  & $6.49\pm0.12$ & $0.2^{+0.1}_{-0.2}$  & $1.61^{+0.07}_{-0.06}$ &
$2.6^{+0.2}_{-0.2}$  & $8.5^{+0.4}_{-0.3}$\\[3pt]
E5 &  &$6.66\pm0.13$ & $0.3^{+0.2}_{-0.2}$  & $1.87^{+0.10}_{-0.09}$ &
$3.2^{+0.3}_{-0.3}$  & $8.3^{+0.3}_{-0.3}$\\[3pt]
A2 &  $4.69\pm0.19$ & & $1.0^{+0.6}_{-0.5}$ & $1.94^{+0.22}_{-0.22}$ &
$7.3^{+1.6}_{-1.4}$ & $18.4^{+1.5}_{-0.7}$\\[3pt]
A3 &  $6.20\pm0.21$ & & $2.0^{+0.5}_{-0.7}$ & $1.92^{+0.17}_{-0.17}$ &
$11.7^{+1.8}_{-2.2}$ & $29.6^{+2.1}_{-1.0}$\\[3pt]
A4 &  $6.23\pm0.21$ & & $1.5^{+0.5}_{-0.6}$ & $1.69^{+0.09}_{-0.16}$ &
$9.7^{+1.9}_{-1.6}$ & $29.3^{+1.2}_{-1.0}$\\[3pt]
A5 &  $5.30\pm0.19$ & & $1.6^{+0.5}_{-0.6}$ & $1.70^{+0.16}_{-0.09}$ &
$8.3^{+1.8}_{-1.3}$ & $24.8^{+2.0}_{-1.4}$\\[3pt]
A6 &  $3.65\pm0.15$ & & $1.1^{+0.7}_{-0.5}$ & $1.79^{+0.22}_{-0.19}$ &
$5.7^{+1.4}_{-1.1}$ & $15.8^{+1.3}_{-0.9}$\\[3pt]
A7 &  $1.32\pm0.10$ & & $0.5^{+0.6}_{-0.5}$  & $1.79^{+0.53}_{-0.43}$ &
$1.8^{+1.1}_{-0.5}$  & $4.8^{+0.5}_{-0.6}$\\
   \noalign{\smallskip}
            \hline
         \end{tabular}
         \label{t:allsources}
\end{table*}


\begin{figure}
\epsfig{figure=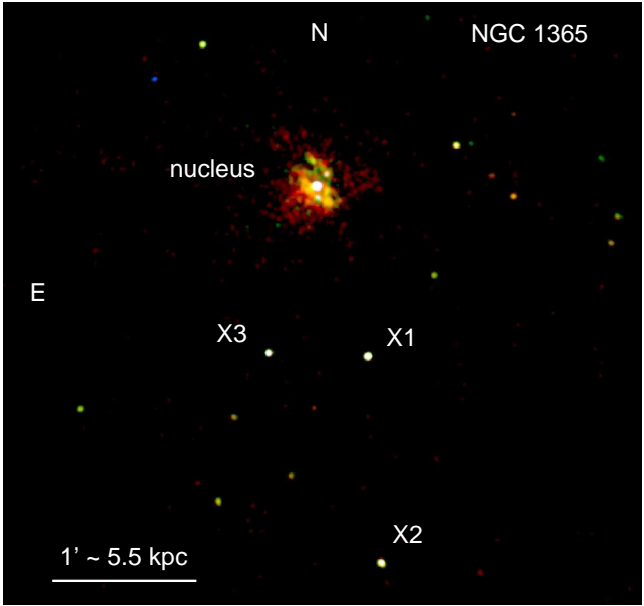, width=8.5cm, angle=0}
\caption{True-colour image from our first {\it Chandra}/ACIS-S 
observation (2002 Dec 24); red corresponds to the $0.3$--$1$ keV band; 
green to $1$--$2$ keV; blue to $2$--$7$ keV. On that occasions, at 
least three sources have emitted X-ray luminosities 
$\ga 3 \times 10^{39}$ erg s$^{-1}$, and are labelled as X1 
(subject of this paper), X2, X3.
The image was smoothed with a Gaussian kernel of radius $1\arcsec$.}
\end{figure}


\begin{figure}
\epsfig{figure=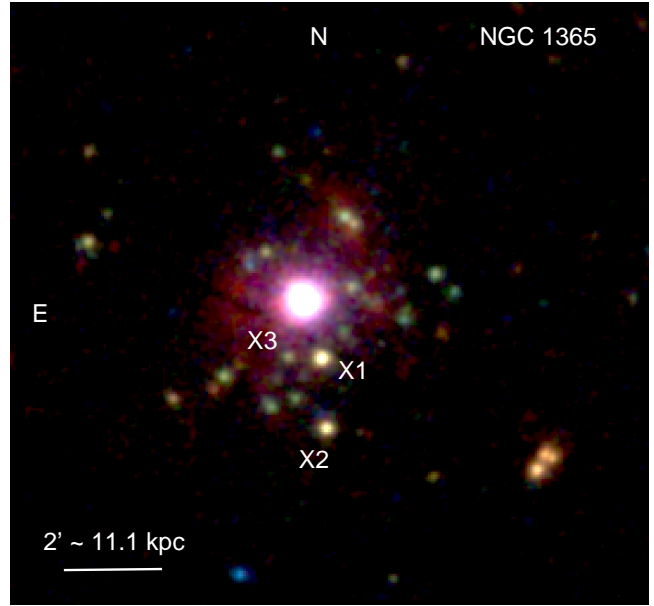, width=8.5cm, angle=0}
\caption{True-colour image from {\it XMM-Newton}/EPIC 
(composite of pn and MOS data from the 2003--2004 observations); 
red corresponds to the $0.3$--$1$ keV band; 
green to $1$--$2$ keV; blue to $2$--$10$ keV. X3 was on average 
much fainter than in the first {\it Chandra} observation (Figure 1).
The image was smoothed with a Gaussian kernel of radius $6\arcsec$.}
\end{figure}



\begin{table*}
      \caption{Summary of the goodness-of-fit ($\chi^2_{\nu}$) of 
simple phenomenological models for each observation, 
and $0.3$--$10$ keV emitted luminosities corresponding 
to each of those fits. Luminosities are in units of $10^{39}$ erg
s$^{-1}$. The results for the two-component models (power-law + diskbb) and 
(power-law $-$ diskbb) are listed only for those spectra 
where they represent an improvement with respect 
to a simple power-law fit. Some spectra (e.g., E2) do not have 
enough counts for a meaningful two-component fit.}
         \begin{tabular}{lccccc}
            \hline
            \noalign{\smallskip}
Observation & Parameter &  \multicolumn{3}{c}{Spectral model} 
& More complex models\\[3pt]
 & & diskbb & power-law & power-law + diskbb &  \\  
            \noalign{\smallskip}
            \hline
            \noalign{\smallskip}
A1 &  $\chi^2_{\nu} =$ & 1.06 (20.1/19) & 0.80 (15.3/19) & 0.66
            (11.3/17) &  See Table 4 \\[3pt]
   &  $L_{0.3-10} =$  & 4.3 & 7.4 & 18.0 &   \\[5pt]
E1 &  $\chi^2_{\nu} =$ & 1.71 (37.7/22) & 0.92 (20.1/22) &  no improv &  \\[3pt]
   &  $L_{0.3-10} =$  & 2.2  & 3.3 & -- &   \\[5pt]
E2 &  $\chi^2_{\nu} =$ & 2.22 (6.7/3) & 2.17 (6.5/3) & no improv &  \\[3pt]
   &  $L_{0.3-10} =$  & 1.7 & 2.2 & -- &   \\[5pt]
E3 &  $\chi^2_{\nu} =$ & 1.14 (13.6/12) & 0.64 (7.71/12) & no improv &  \\[3pt]
   &  $L_{0.3-10} =$  & 1.7 & 2.9 & -- &   \\[5pt]
E4 &  $\chi^2_{\nu} =$ & no fit & 1.15 (158.2/138) & 0.95 (129.1/136)
   & See Table 5 \\[3pt]
   &  $L_{0.3-10} =$  & no fit & 8.5 & 10.5 &   \\[5pt]
E5 &  $\chi^2_{\nu} =$ & no fit & 1.33 (155.9/117) & 1.02 (117.7/115)
   & See Table 6 \\[3pt]
   &  $L_{0.3-10} =$  & no fit & 8.3 & 9.2 &   \\[5pt]
A2 &  $\chi^2_{\nu} =$ & 1.15 (39.4/34) & 0.99 (33.7/34) & no improv &  \\[3pt]
   &  $L_{0.3-10} =$  & 11.7 & 18.4 & --  &   \\[5pt]
A3 &  $\chi^2_{\nu} =$ & 0.82 (39.5/48) & 0.64 (30.6/48) & no improv &
   See Table 7 \\[3pt]
   &  $L_{0.3-10} =$  & 18.9 & 29.6 & -- &  \\[5pt]
A4 &  $\chi^2_{\nu} =$ & 0.80 (39.1/49) & 0.81 (39.5/49) & no improv &
   See Table 7 \\[3pt]
   &  $L_{0.3-10} =$  & 20.1 & 29.3 & -- &  \\[5pt]
A5 &  $\chi^2_{\nu} =$ & 0.71 (31.2/44) & 0.75 (32.9/44) & no improv &
   See Table 7\\[3pt]
   &  $L_{0.3-10} =$  & 17.4 & 24.8  & -- &   \\[5pt]
A6 &  $\chi^2_{\nu} =$ & 0.50 (15.5/31) & 0.38 (11.7/31) & no improv &
   \\[3pt]
   &  $L_{0.3-10} =$  & 10.6 & 15.8 & -- &   \\[5pt]
A7 &  $\chi^2_{\nu} =$ & 1.28 (14.1/11) & 0.90 (9.88/11) & no improv &  \\[3pt]
   &  $L_{0.3-10} =$  & 2.8 & 4.8 & -- &   \\
   \noalign{\smallskip}
            \hline
         \end{tabular}
         \label{t:allsources}
\end{table*}


\begin{figure}
\epsfig{figure=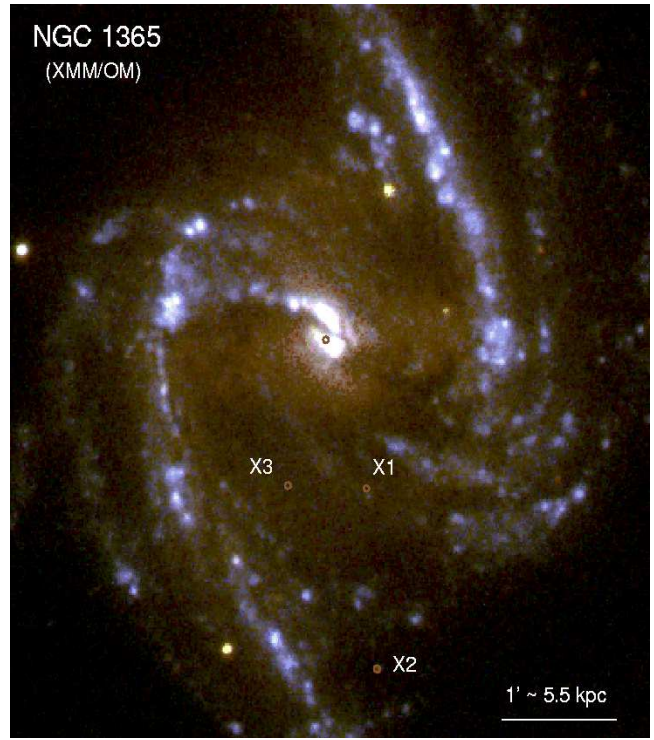, width=8.5cm, angle=0}
\caption{True-colour UV image of NGC\,1365 
from {\it XMM-Newton}/OM (observed on 2004 Jan 17). Red 
corresponds to the $U$ filter, green to the $UVW1$ filter, 
and blue to the $UVM2$ filter. The positions of the nucleus 
and of the three ULX detected in the {\it Chandra} field 
(Figure 1) are overplotted as red circles, with a radius 
of $1\farcs5$, which is a conservative estimate 
of the uncertainty. North is up, East to the left.} 
\end{figure}

\begin{figure}
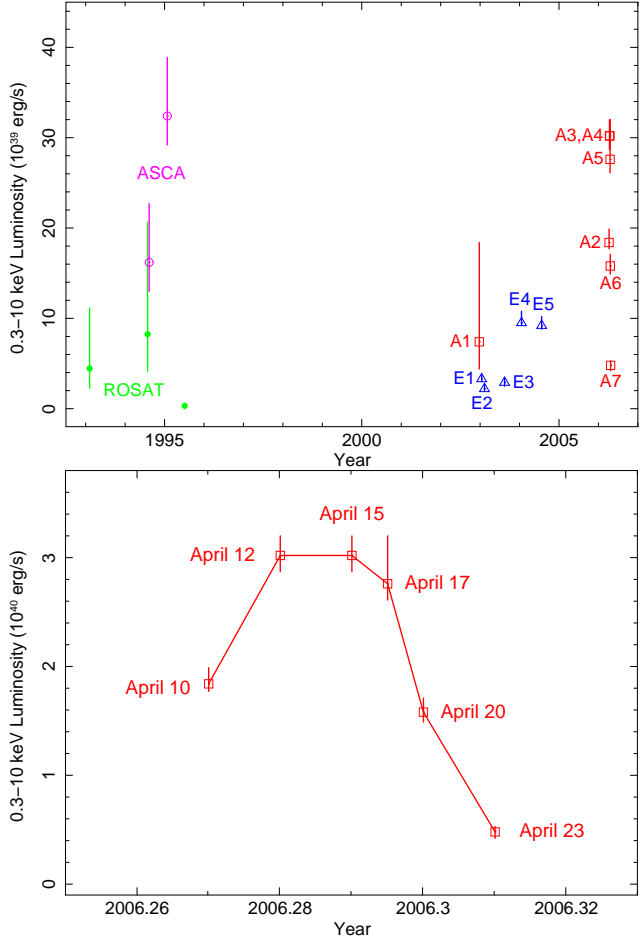

\epsfig{figure=lightcurve_new1.ps, width=6.2cm, angle=270}\\
\epsfig{figure=lightcurve_new2.ps, width=6.2cm, angle=270}
\caption{Top panel: historical lightcurve of NGC\,1365 X1. Unabsorbed 
fluxes for the {\it ROSAT} (green filled circles) observations 
have been estimated from the instrumental count rates published 
in the literature, assuming a power-law photon index $\Gamma = 1.7$ 
and Galactic absorption. We estimated the error bars by varying the spectral 
slope between $1.6$ and $2.1$, and increasing the column density to $2
\times 10^{21}$ cm$^{-2}$. For the {\it ASCA} observations  
(magenta open circles), we have re-extracted and analysed 
the data. For the {\it Chandra} (red squares) 
and {\it XMM-Newton} (blue triangles) observations, we used 
the best-fitting models listed in Tables 3--7. 
We converted fluxes to luminosities assuming isotropic emission 
and a distance of $19$ Mpc. Bottom panel: zoomed-in view of the luminosity 
variations over the timespan of our {\it Chandra} observations.}
\end{figure}

\begin{figure}
\epsfig{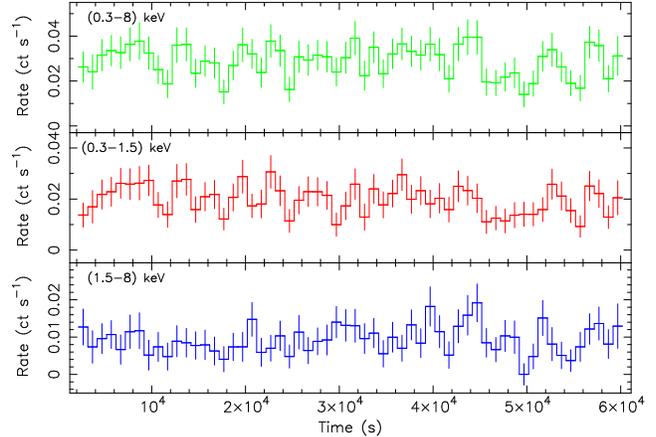}
\caption{{\it XMM-Newton}/EPIC-pn lightcurves during observation E4, 
in the total ($0.3$--$8$ keV), soft ($0.3$--$1.5$ keV) 
and hard ($1.5$--$8$ keV) bands. 
Short-term variability is marginally detected in the total band.}
\end{figure}

\begin{center}
   \begin{table}
      \caption{Best-fitting parameters 
to the {\it Chandra}/ACIS spectrum of X1 
from observaton A1. The {\small XSPEC} model is
{\tt tbabs}$_{\rm Gal}~\times$ {\tt tbabs} $\times$ ({\tt diskbb}+{\tt po}). 
The quoted errors are the 90\% confidence limit and
$N_{\rm H,Gal} = 1.4 \times 10^{20}$ cm$^{-2}$ (Dickey \& Lockman 1990).
The power-law normalization $K_{\rm po}$ is in units 
of $10^{-5}$ photons keV$^{-1}$ cm$^{-2}$ s$^{-1}$ at 1 keV.}
         \label{table1c}
\begin{center}
         \begin{tabular}{lr}
            \hline
            \noalign{\smallskip}
            Parameter    & Value \\[2pt]
            \noalign{\smallskip}
            \hline
            \noalign{\smallskip}
            \noalign{\smallskip}
                $N_{\rm H}~(10^{21}~{\rm cm}^{-2})$ 
	                & $4.9^{+4.1}_{-2.9}$ \\[3pt]
                $kT_{\rm in}$~(keV) & $0.17^{+0.18}_{-0.06}$ \\[3pt]
                $K_{\rm dbb}$ & $25.9^{+211.0}_{-25.1}$ \\[3pt]
                $\Gamma$  & $1.83^{+0.60}_{-0.60}$\\[3pt]
                $K_{\rm po}~(10^{-5})$ 
                        & $2.5^{+2.7}_{-1.5}$\\
            \noalign{\smallskip}
            \hline
            \noalign{\smallskip}
                $\chi^2_\nu$ & $0.66 (11.3/17)$ \\[3pt] 
                $L^{\rm dbb}_{0.3{\rm -}10}~(10^{39}~{\rm erg~s}^{-1})$ 
                        & $10.6^{+21.5}_{-8.0}$\\[3pt]
                $L^{\rm tot}_{0.3{\rm -}10}~(10^{39}~{\rm erg~s}^{-1})$ 
                        & $18.0^{+5.5}_{-9.5}$\\
            \noalign{\smallskip}
            \hline
         \end{tabular}
\end{center}
   \end{table}
\end{center}

\begin{center}
   \begin{table}
      \caption{Best-fitting parameters 
to the combined {\it XMM-Newton}/EPIC spectrum of X1 
from observaton E4. The {\small XSPEC} models are: 
model 1 = {\tt tbabs}$_{\rm Gal}~\times$ {\tt tbabs} $\times$ ({\tt diskbb}+{\tt po}); 
model 2 = {\tt tbabs}$_{\rm Gal}~\times$ {\tt tbabs} $\times$ ({\tt diskbb}+{\tt bknpo})
The quoted errors are the 90\% confidence limit and
$N_{\rm H,Gal} = 1.4 \times 10^{20}$ cm$^{-2}$ (Dickey \& Lockman 1990).} 
         \label{table1c}
\begin{center}
         \begin{tabular}{lcc}
            \hline
            \noalign{\smallskip}        
            Parameter    & Model 1 value   & Model 2 value \\[2pt]
            \noalign{\smallskip}
            \hline
            \noalign{\smallskip}
            \noalign{\smallskip}
                $N_{\rm H}~(10^{21}~{\rm cm}^{-2})$ 
	                & $1.0^{+0.6}_{-0.5}$ &$0.7^{+0.7}_{-0.5}$\\[3pt]
                $kT_{\rm in}$~(keV) & $0.23^{+0.03}_{-0.05}$ &$0.27^{+0.10}_{-0.07}$\\[3pt]
                $K_{\rm dbb}$ & $1.46^{+0.16}_{-0.23}$ &$0.67^{+3.20}_{-0.52}$\\[3pt]
                $\Gamma_1$  & $1.37^{+0.14}_{-0.08}$ &$1.13^{+0.20}_{-0.34}$\\[3pt]
                $E_{\rm br}$  & -- &$5.7^{+0.8}_{-1.2}$\\[3pt]
                $\Gamma_2$  & -- &  $2.47^{+1.57}_{-0.78}$\\[3pt]
                $K_{\rm po}~(10^{-5})$ 
                        & $1.9^{+0.5}_{-0.3}$ & $1.5^{+0.5}_{-0.5}$\\
            \noalign{\smallskip}
            \hline
            \noalign{\smallskip}
                $\chi^2_\nu$ & $0.95 (129.1/136)$ & $0.92 (122.8/134)$\\[3pt] 
                $L^{\rm dbb}_{0.3{\rm -}10}~(10^{39}~{\rm erg~s}^{-1})$ 
                        & $2.4^{+1.6}_{-0.7}$ & $2.3^{+2.6}_{-0.6}$\\[3pt]
                $L^{\rm tot}_{0.3{\rm -}10}~(10^{39}~{\rm erg~s}^{-1})$ 
                        & $10.5^{+0.5}_{-1.1}$ & $9.5^{+1.3}_{-0.3}$\\
            \noalign{\smallskip}
            \hline
         \end{tabular}
\end{center}
   \end{table}
\end{center}

\begin{center}
   \begin{table}
      \caption{Best-fitting parameters 
to the combined {\it XMM-Newton}/EPIC spectrum of X1 
from observaton E5. The {\small XSPEC} model is
{\tt tbabs}$_{\rm Gal}~\times$ {\tt tbabs} $\times$ ({\tt diskbb}+{\tt po}). 
The quoted errors are the 90\% confidence limit and
$N_{\rm H,Gal} = 1.4 \times 10^{20}$ cm$^{-2}$ (Dickey \& Lockman 1990).} 
         \label{table1c}
\begin{center}
         \begin{tabular}{lr}
            \hline
            \noalign{\smallskip}
            Parameter    & Value \\[2pt]
            \noalign{\smallskip}
            \hline
            \noalign{\smallskip}
            \noalign{\smallskip}
                $N_{\rm H}~(10^{21}~{\rm cm}^{-2})$ 
	                & $<0.2$ \\[3pt]
                $kT_{\rm in}$~(keV) & $0.41^{+0.05}_{-0.05}$ \\[3pt]
                $K_{\rm dbb}$ & $0.13^{+0.01}_{-0.04}$ \\[3pt]
                $\Gamma$  & $0.82^{+0.23}_{-0.39}$\\[3pt]
                $K_{\rm po}~(10^{-5})$ 
                        & $0.7^{+0.4}_{-0.3}$\\
            \noalign{\smallskip}
            \hline
            \noalign{\smallskip}
                $\chi^2_\nu$ & $1.02 (117.7/115)$ \\[3pt] 
                $L^{\rm dbb}_{0.3{\rm -}10}~(10^{39}~{\rm erg~s}^{-1})$ 
                        & $2.9^{+0.5}_{-0.4}$\\[3pt]
                $L^{\rm tot}_{0.3{\rm -}10}~(10^{39}~{\rm erg~s}^{-1})$ 
                        & $9.2^{+1.0}_{-0.3}$\\
            \noalign{\smallskip}
            \hline
         \end{tabular}
\end{center}
   \end{table}
\end{center}

\begin{figure}
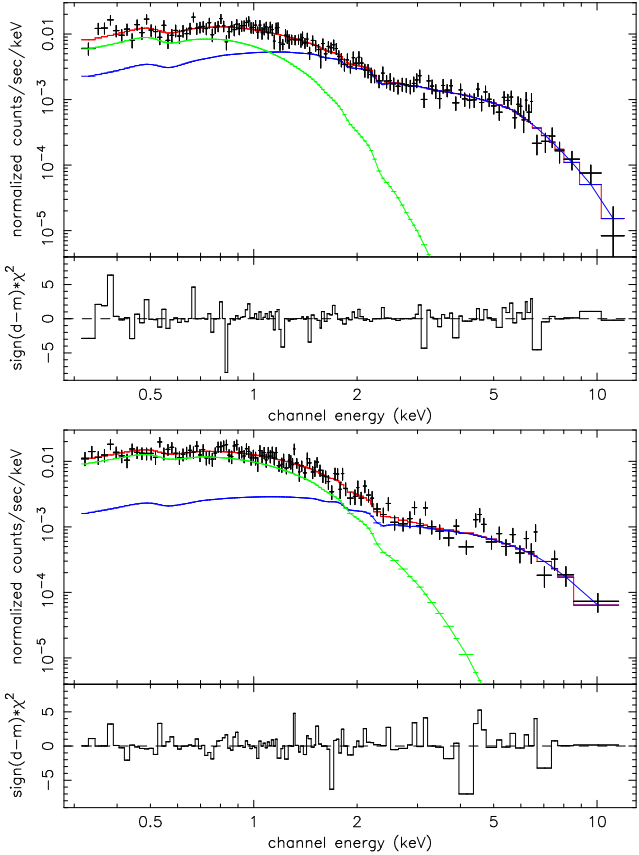

\epsfig{figure=epic4_spectrum.ps, width=5.65cm, angle=270}\\
\epsfig{figure=epic5_spectrum.ps, width=5.65cm, angle=270}
\caption{Top panel: combined {\it XMM-Newton}/EPIC spectrum and $\chi^2$ 
residuals from observation E4, 
fitted with an absorbed broken power-law plus disk-blackbody component. 
The best-fitting parameters are given in Table 5.
Bottom panel: the same for observation E5; the best-fitting parameters 
are given in Table 6.}
\end{figure}


\begin{figure}
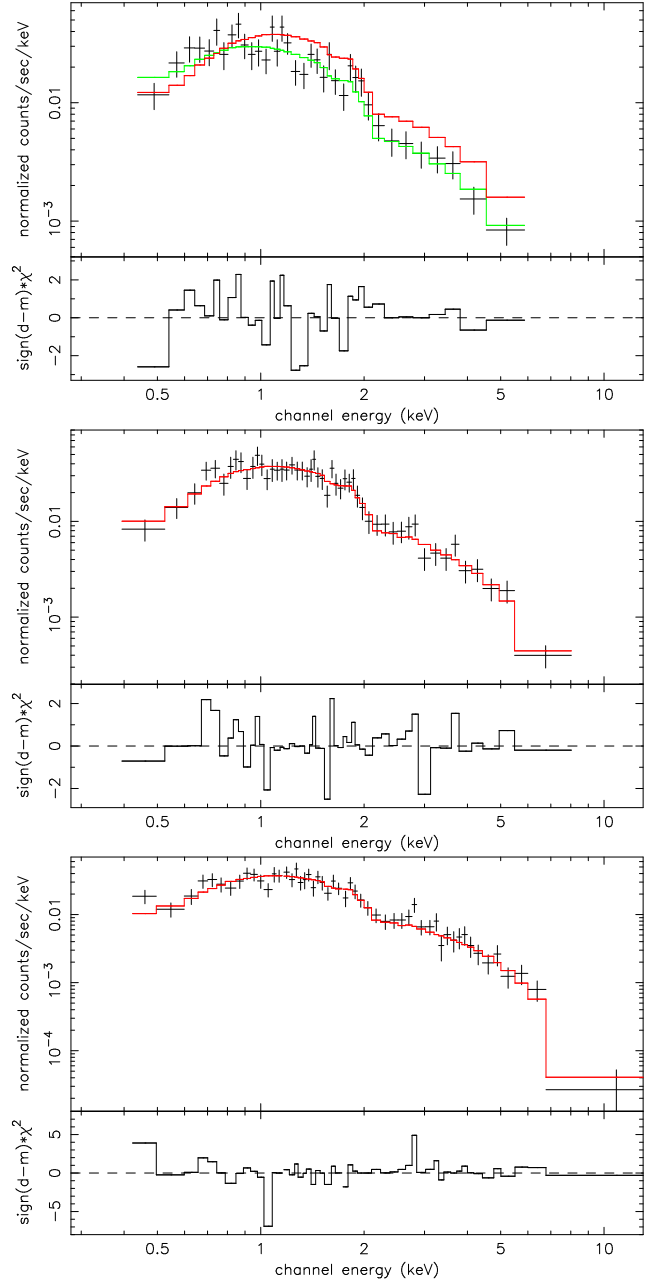

\epsfig{figure=acisa_spectrum.ps, width=5.65cm, angle=270}\\
\epsfig{figure=acisb_spectrum.ps, width=5.65cm, angle=270}\\
\epsfig{figure=acisc_spectrum.ps, width=5.65cm, angle=270}
\caption{Top panel: {\it Chandra}/ACIS spectrum and $\chi^2$ 
residuals from observation A2. 
The best-fitting absorbed 
power-law model is plotted in green; fit parameters are listed 
in Table 2. For comparison, the best-fit model from 
the outburst peak (observation A3) is overplotted in red.
Middle panel: the same for observation A3 (peak of the outburst).
Bottom panel: the same for observation A4 (identical to A3).}
\end{figure}



\begin{figure}
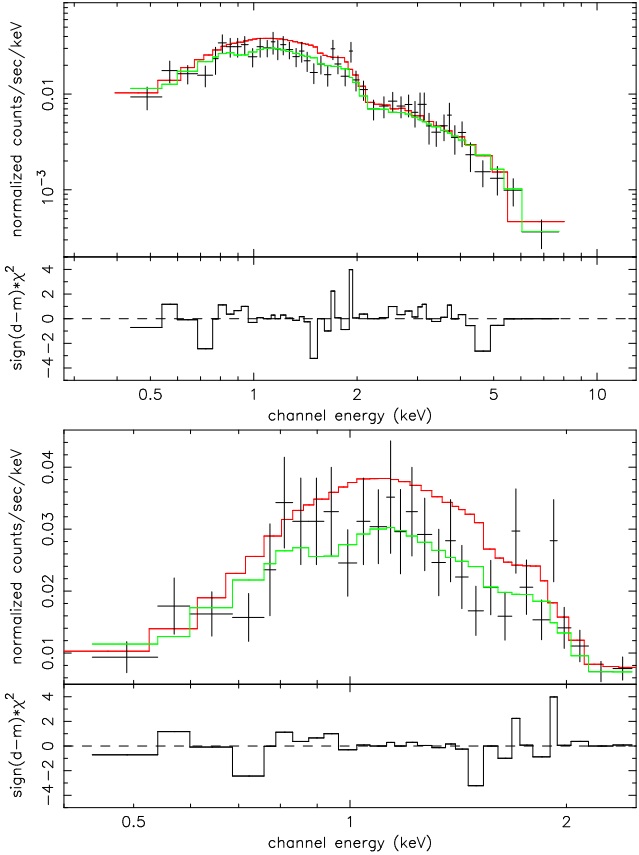

\epsfig{figure=acis_absori2.ps, width=5.65cm, angle=270}
\epsfig{figure=acis_absori1.ps, width=5.65cm, angle=270}
\caption{Top panel: {\it Chandra}/ACIS spectrum and $\chi^2$ 
residuals from observation A5, fitted with a power-law 
absorbed by both neutral and ionized gas. The best-fitting 
model is overplotted in green, and its parameters are given in Table 7. 
Bottom panel: expanded view, in linear scale, of the soft X-ray 
spectral data and model (plotted in green) for A5, compared 
with the (identical) best-fit model for the A3 and A4 spectra 
(plotted in red), to emphasize the effect of the ionized 
absorption in A5.}
\end{figure}

\begin{figure}
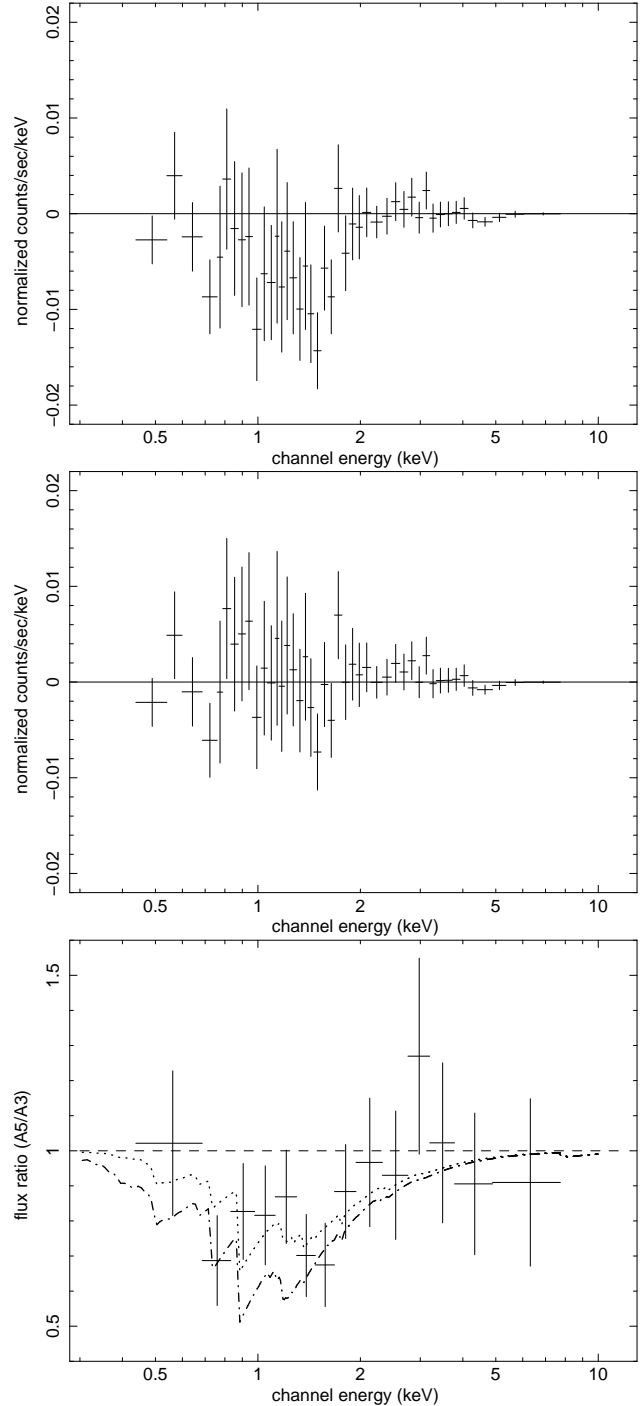

\epsfig{figure=residuals_plot.ps, width=6.2cm, angle=270}
\epsfig{figure=residuals_plot2.ps, width=6.2cm, angle=270}
\epsfig{figure=ratio_plot2.ps, width=6.2cm, angle=270}
\caption{Top panel: spectral residuals from observation A5, 
fitted with a power-law absorbed by neutral gas, with 
same fit parameters as A3 and A4. Middle panel: spectral residuals 
when an additional ionized absorption component is added to the model.
Bottom panel: model-independent ratio of the observed fluxes 
from A5 and A3, showing again the deficit of soft X-ray photons 
in A5. Dotted line: predicted ratio for an ionized absorber 
with $N_{\rm H} = 10^{22}$ cm$^{-2}$ and $\xi = 100$. 
Dash-dotted line: predicted ratio for $N_{\rm H} = 10^{22}$ cm$^{-2}$ 
and $\xi = 50$.}
\end{figure}

\subsection{Thermal and non-thermal spectral components}

In all the observations with a sufficient number of counts, 
the X-ray spectrum appears dominated by a broad 
power-law-like component. In particular, the wider spectral 
coverage of {\it XMM-Newton} clearly rules out 
the possibility that the source is in a thermal dominant 
state, when most of the emission comes from 
a standard multicolour disk. Most ULXs 
(especially those with X-ray luminosities $\ga 10^{40}$ erg s$^{-1}$) 
seem to share this phenomenological property, 
indicating perhaps that most of the disk photons 
are Compton-upscattered.

The two longest {\it XMM-Newton} observations (E4 and E5) 
have a sufficiently high signal-to-noise ratio, allowing 
a more detailed spectral analysis.
Firstly, in both datasets, significant systematic residuals 
at low energies are found when the source spectra are fitted 
with a simple power-law model: an additional soft component 
is required\footnote{Simple power-law fits yield  
$\chi^2_\nu = 158/138$ and $156/117$ for E4 and E5 respectively 
(Table 3), as opposed to $\chi^2_\nu = 129/136$ and $118/115$ 
for disk-blackbody plus power-law fits. To verify 
that this is a statistically significant improvement, 
we have simulated pure power-law spectra using {\it fakeit} 
in {\small XSPEC}, and fitted them both with power-law 
and disk-blackbody plus power-law models; we have then 
compared the decrease in $\chi^2_\nu$ with what we obtain 
from the real datasets. This method 
is essentially equivalent to an F-test (Protassov 
et al.~2002).}.
The soft excess contributes
$\approx 1/4$ of the emitted luminosity\footnote{Throughout 
this paper, for simplicity, luminosities are defined  
as $4\pi d^2$ times the respective fluxes, assuming isotropic 
emission for both the power-law and the disk-blackbody components. 
More rigorously, disk-blackbody luminosities should be 
calculated as $(2\pi d^2/\cos i)$ times the flux, where $\cos i$ 
is the (unknown) viewing angle to the disk. The difference 
between the two estimates is less than a factor of 2 
for $i \la 75^{\circ}$.} in the $0.3$--$10$ keV
band, but dominates below $\sim 1.5$ keV 
(Figure 6 and Tables 5, 6). 
This emission is well fitted by a standard disk-blackbody 
model, with colour temperatures $\sim 0.3$--$0.4$ keV. 
Such temperatures are intermediate between those generally 
found in ``typical'' ULXs (Miller, Fabian \& Miller~2004; 
Stobbart et al.~2006) 
and those observed from the disks of Galactic stellar-mass BHs 
(e.g., Remillard \& McClintock 2006). 
A possible soft excess is marginally found also for the  
{\it Chandra} observation A1 (Table 4), but there are not 
enough counts to significantly constrain its flux and temperature. 

Secondly, we find that in E4 there is a significant steepening of the continuum 
slope above $\approx 5.7$ keV (Figure 6). Similar breaks are 
also found in various other bright ULXs: they have been interpreted 
as the effect of Comptonization in a warm ($kT_e \la 5$ keV), moderately 
optically-thick ($\tau \sim 2$--$4$) corona (Stobbart et al.~2006; 
Goad et al.~2006; Dewangan, Griffiths \& Rao 2006).
The alternative interpretation  
for the spectral steepening is that the harder emission 
above $\sim 1.5$ keV is in fact from a hot ($kT_{\rm in} \sim 2$ keV) 
accretion disk, so that the steeper continuum above $\sim 6$ keV would  
represent the Wien part of the thermal disk component, 
somewhat modified (broadened) by Comptonization. 
Physically, this may be consistent with a ``slim disk'' 
model, also sometimes used to explain some ULX spectra 
(e.g., Ebisawa et al.~2003; Mizuno et al.~2007). 
Although such model provides an equally good fit 
for the E4 spectrum, it is not acceptable for E5 
nor for the brightest {\it Chandra} spectra; 
therefore we do not consider it as a viable physical 
scenario for this particular source.

Thirdly, we find that the true value of the power-law 
photon index may be lower (harder spectrum) when the thermal 
component is also taken into account. 
Simple power-law fits to the E4 and E5 spectra 
give $\Gamma \approx 1.6$--$1.7$ (Table 2). 
But when the same E4 spectrum is fitted
with a broken power-law plus disk-blackbody,
the photon index is $\Gamma \approx 1.1$ 
below the break, steepening to  $\Gamma \approx 2.5$ 
at higher energies (Table 5). 
For E5, the photon index is $\Gamma \approx 0.8$ (Table 6), 
at least up to energies $\approx 10$ keV; we cannot determine 
where the expected spectral break occurred on that occasion.
We tried to find evidence of high-energy spectral breaks 
also in our series of {\it Chandra} spectra. However, 
the ACIS sensitivity drops rapidly above $6$ keV, making 
this task impossible for such short observations.

\begin{figure}
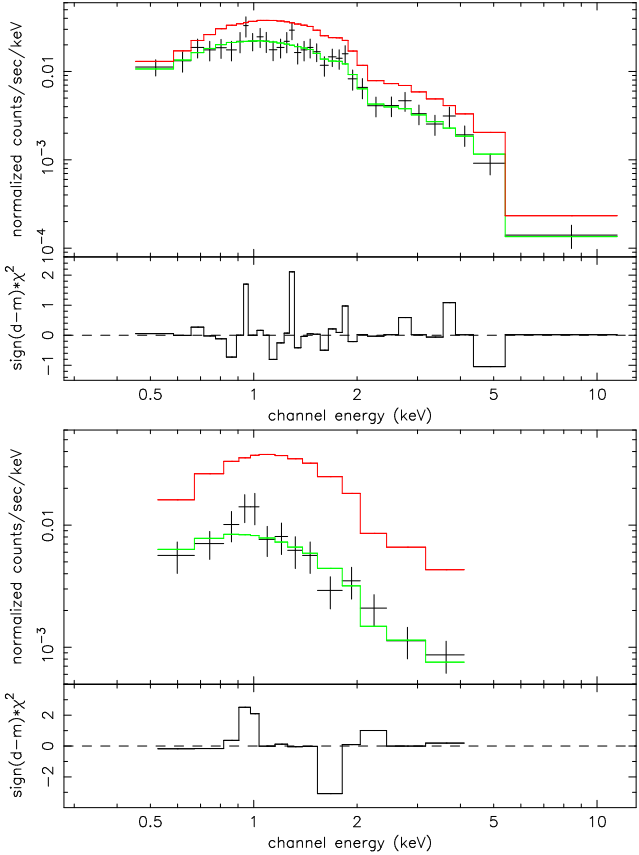

\epsfig{figure=acise_spectrum.ps, width=5.65cm, angle=270}\\
\epsfig{figure=acisf_spectrum.ps, width=5.65cm, angle=270}
\caption{Top panel: {\it Chandra}/ACIS spectrum and $\chi^2$ 
residuals from observation A6, fitted with an absorbed power-law. 
Bottom panel: the same for observation A7.
For both spectra, the best-fitting parameters are listed in Table 2. 
The model corresponding to the peak of the outburst (A3)
is plotted in red, for comparison.}
\end{figure}


\subsection{The 2006 outburst sequence}

The most important result from the sequence of {\it Chandra} 
observations in 2006 April is that we caught the rise, peak 
and decline of an outburst, in which the source 
reached unabsorbed X-ray luminosities 
$\approx 3 \times 10^{40}$ erg s$^{-1}$, 
similar to the values observed by {\it ASCA} in 
the 1995 January outburst. It was considered remarkable 
at that time that the source must have risen and declined 
by more that an order of magnitude in less than a few weeks. 
In fact, the 2006 outburst shows that the characteristic 
timescale for the rise and decline is only a few days 
(Figure 4 and Tables 1, 2). The full sequence of spectra, fitted with 
absorbed power-law models, is shown in Figures 7--10. 

The outburst peaked between observations A3 (2006 April 12) 
and A4 (2006 April 15). The best-fitting models 
are consistent with being the same for both spectra, 
within the errors: a simple power-law component 
($\Gamma = 1.9 \pm 0.2$ for A3, $\Gamma = 1.7 \pm 0.1$ for A4) 
with neutral absorption, and essentially identical flux. 
In fact, the A5 spectrum 
is also essentially identical to A3 and A4 below $\approx 0.7$ keV 
and above $\approx 2$ keV. The difference, however, 
is that the A5 spectrum has a statistically significant 
``soft deficit'' at energies $\approx 0.7$--$2$ keV (Figures 8, 9), 
that cannot be modelled with additional cold absorption 
or with any kind of thermal emission.
This feature suggests the presence of additional 
ionized absorption in the A5 spectrum.

To test this hypothesis, we fitted those three outburst 
spectra simultaneously, assuming the same primary spectrum 
(same power-law slope and normalization) and the same amount 
of neutral absorption, leaving only the amount of ionized
absorption ({\tt absori} model in {\footnotesize XSPEC}) 
free to vary between them. As expected, 
the ionized column densities in A3 and A4 is consistent with $0$; 
we then fixed both of them at $0$ to obtain a better constraint on 
the error range for the other parameters; we also fixed 
the metal abundance to the solar value. The ionized column 
density of the A5 spectrum is $\approx 10^{22}$ cm$^{-2}$, 
with a corresponding ionization parameter 
$\xi = L_{>13.6 {\rm eV}}/nr^2 \approx 100$, 
or $U \approx 0.5$--$1$, if we adopt the definition 
of ionization parameter used for example 
by Murray et al.~(1995).
We also note that the neutral absorbing column density around 
the peak of the outburst (A3, A4, A5) appears to be higher 
($\approx 1.5$--$2 \times 10^{21}$ cm$^{-2}$) than the value found 
at other epochs, when it is typically 
$\approx 5 \times 10^{20}$ cm$^{-2}$.

The A6 spectrum, taken three days after A5, already shows 
the beginning of the decline, with a reduction 
of the emitted luminosity by a factor $\approx 2$ 
(Figure 10, top panel). The spectrum is well fitted with a simple 
power law absorbed by neutral gas. After three more days 
(observation A7: Figure 10, bottom panel) the luminosity had declined 
by another factor of $\approx 3.5$.
In summary, the X-ray luminosity has an $e$-folding timescale 
$\approx 3$ days in the decline. 

We could compare this timescale with typical decline timescales 
observed at the end of Galactic stellar-mass BH outbursts. 
For example, the $e$-folding timescale for XTE J1550$-$564 
at the end of the 1998 outburst was 11 days (Wu et al.~2002); 
and it was $\approx 2$ weeks for GRO J1655$-$40 
at the end of the 1996--1997 outburst 
(Sobczak et al.~1999). Those decline timescales 
very likely correspond to the viscous timescale 
on which the disk is emptied after the mass inflow rate 
from the donor star is strongly reduced; they depend 
on the size of the Roche lobe and on the effective viscosity. 
As the disk is emptied, a spectral state transition occurs 
from the high/soft to the low/hard state.
However, the situation for NGC\,1365 X1 is entirely 
different. In our interpretation, the X-ray spectrum 
is power-law dominated during the outburst; this is consistent 
with the suggestion that ULXs are in a state similar 
to the very high state of stellar-mass BHs 
(Roberts et al.~2006). 
Hence, we argue that the flare observed in the {\it Chandra} 
observations is probably comparable to the hard flares seen 
in the very high state (often at the very beginning of an outburst) 
of stellar-mass BHs (for example, 
XTE J1550$-$564: Sobczak et al.~2000; 
GRO J1655$-$40: Sobczak et al.~1999;
XTE J1859$+$226: Brocksopp et al.~2002;
4U 1630$-$47: Trudolyubov, Borozdin \& Priedhorsky 2001). 
Typical amplitudes of those X-ray flares are a factor 
of $\approx 2$--$5$; typical decline timescales 
are $\approx 2$--$4$ days.
Physically, such flares may correspond to 
the formation and ejection of a Comptonizing 
region in the inner part of the accretion flow.
This is consistent with the decline being 
associated with an outflow.

Finally, we checked whether there is any evidence of a soft excess 
in the 2006 outburst spectra, that can be attributed to disk 
emission. None of the individual spectra are improved 
by adding a disk-blackbody component. We find the tightest constraint 
by modelling the coadded spectrum of the observations A3 and A4.
Any disk-blackbody component with inner-disk colour temperature 
$kT_{\rm in} = 0.20$ keV must have an emitted luminosity 
$< 3.0 \times 10^{39}$ erg s$^{-1}$ ($90\%$ confidence level) 
in the $0.3$--$10$ keV band. If the inner-disk colour temperature 
$kT_{\rm in} = 0.40$ keV, the upper limit to the emitted luminosity 
is $2.0 \times 10^{39}$ erg s$^{-1}$ in the same band.
Those apper limits are comparable to the inferred disk fluxes 
from E4 and E5. We cannot tell whether the disk has significantly 
decreased in flux during the outburst, but clearly it did 
not become more luminous. We argue that the disk emission represented 
$\sim 1/4$ of the X-ray emission when the total $0.3$--$10$ keV 
luminosity was $\approx 1 \times 10^{40}$ erg s$^{-1}$; 
and it represented $< 10\%$ of the X-ray emission when 
the source had brightened to $\approx 3 \times 10^{40}$ erg s$^{-1}$.
Thus, the brightening is entirely characterized by the enhancement 
of the power-law-like component.

\subsection{Evidence of outflows}

We further examined the behaviour at the peak of the 2006 outburst, 
when spectral fitting suggested the onset of an outflow. 
We directly compared the observed spectra A3 and A5, 
after rebinning them to increase the signal-to-noise. 
The ratio of the two spectra confirms in a model-independent way 
that a portion of the soft X-ray spectrum is absorbed in A5, 
consistent with an ionized column density 
$N \approx 10^{22}$ cm$^{-2}$, and an 
ionization parameter $\xi \approx 100$ (Figure 9, bottom panel).

Those two characteristic values, similar to those often 
found in Seyfert galaxies (e.g., Crenshaw, Kraemer \& George 2003), 
are clearly inconsistent with a simple uniform shell 
of absorbing material around the X-ray source.
The ionizing luminosity (above $13.6$ eV) emitted by the ULX is 
$\approx (5\pm1) \times 10^{40}$ erg s$^{-1}$. This would 
imply a characteristic size $\approx 5 \times 10^{16}$ cm 
for the absorbing gas, 
much larger than the characteristic size of the binary system. 
Considering that the increase in the absorption column density 
has occurred over only few days, and that typical maximum radial 
velocities of radiatively-driven outflows are $\sim 10^9$ cm 
s$^{-1}$, we cannot expect the outflow to have reached 
distances larger than a few $\times 10^{14}$ cm. 

For a more accurate study, we tried using 
a radiatively-driven disk-wind model (Murray \& Chiang 1995; 
Murray et al.~1995). We assume that $L \sim L_{\rm Edd}$, 
in which case we expect a mass loss rate in the wind 
$\dot{M}_{\rm w} \sim (10^{27}/v_{\rm in})\,(M/100 M_{\odot})$ 
g s$^{-1}$. If the wind is launched from the inner disk, 
the initial (thermal) velocity $v_{\rm in} \approx 0.5$--$1 \times 10^7$ 
cm s$^{-1}$ and the mass loss rate $\sim 10^{20}$ g s$^{-1}$. 
The column density, integrated along a streamline, is
\begin{equation}
N \approx \frac{\dot{M}_{\rm w} \ln(v_{\infty}/v_{\rm in})}
{4\pi r_{\rm in} m_{\rm p} v_{\infty}} 
\end{equation}
(Murray \& Chiang 1995).
For typical $v_{\infty} \sim 10^9$ cm s$^{-1}$ 
and a column density $\approx 10^{22}$ cm$^{-2}$ 
as inferred from our spectral modelling, 
the wind would have to be launched from radii 
$r_{\rm in} \sim 10^{12}$ cm $\sim 10^5$ gravitational 
radii for a $100 M_{\odot}$ BH. This is two orders 
of magnitude larger than expected, by comparison with 
disk winds in AGN. 

In summary, the problem we face is that if we were 
looking into a disk wind launched from 
$r_{\rm in} \la 10^{10}$ cm we would see 
a column density $N \ga 10^{24}$ cm$^{-2}$ 
and, for $L \approx 5 \times 10^{40}$ erg s$^{-1}$, 
an ionization parameter a few orders of magnitude higher 
than inferred from the soft X-ray absorption, and also too high 
for the wind to be radiatively accelerated. 
Increasing the mass outflow 
rate would decrease the ionization state of the gas 
but increase the column density. An analogous problem 
was discussed by Schurch \& Done (2006). 
A possible way to keep 
the ionization parameter in the wind sufficiently low 
is to assume (following Murray et al.~1995, 
Proga, Stone \& Kallman 2000, Proga \& Kallman 2004) 
that the outflowing gas does not see most of the soft X-ray photons: 
it is shielded by ``hitchhiking gas'' at the inner face 
of the wind, or by the disk atmosphere at the base of the wind.
In particular, this is expected to occur when the mass outflow 
rate is $\ga$ Eddington accretion rate, because 
the outflow is Compton thick (Pounds et al.~2003) 
and the gas in the outer wind does not directly see 
the soft X-ray irradiation. However, we do see 
most of the X-ray flux from X1: the ionized absorber 
is only removing $\approx 10$\% of the $0.3$--$2$ keV 
flux from {\it our} line of sight (Figure 11). 
Qualitatively, we can explain this by assuming a low 
filling factor for the absorber ($\la 10$\%). 
Alternatively, we can speculate that the observed 
soft X-ray flux has a direct (unabsorbed) component 
and an additional ($\approx 10$\%) component 
scattered into our line of sight by the outflowing gas; 
it is only the latter that bears the imprint 
of ionized absorption.

\begin{center}
   \begin{table}
      \caption{Best-fitting parameters 
to the {\it Chandra}/ACIS spectra of X1 
from observation A3, A4 and A5, at the peak of the 2006 April 
outburst. The {\small XSPEC} model is
{\tt tbabs}$_{\rm Gal}~\times$ {\tt tbabs} $\times$ 
{\tt absori} $\times$ {\tt po}. 
The quoted errors are the 90\% confidence limit and
$N_{\rm H,Gal} = 1.4 \times 10^{20}$ cm$^{-2}$ (Dickey \& Lockman 1990).} 
         \label{table1c}
\begin{center}
         \begin{tabular}{lr}
            \hline
            \noalign{\smallskip}
            Parameter    & Value \\[2pt]
            \noalign{\smallskip}
            \hline
            \noalign{\smallskip}
            \noalign{\smallskip}
                $N_{\rm H}~(10^{21}~{\rm cm}^{-2})$ 
	                & $0.19^{+0.04}_{-0.03}$ \\[3pt]
                $\Gamma$  & $1.88^{+0.11}_{-0.11}$\\[3pt]
                $K_{\rm po}~(10^{-5})$ 
                        & $11.5^{+1.7}_{-1.4}$\\[3pt]
                $T^{\rm abs}$~(K)  & $10^5$ (fixed)\\[3pt]
                $Z^{\rm abs}$~($Z_{\odot}$)  & $1$ (fixed)\\[3pt]
                $\xi^{\rm abs}$~(CGS)  & $111^{+184}_{-83}$ \\[3pt]
                $N_{\rm H}^{\rm abs}~(10^{22}~{\rm cm}^{-2})$ 
	                & for A3: $0$ \\[3pt]
	                & for A4: $0.2^{+0.8}_{-0.2}$ \\[3pt]
	                & for A5: $1.0^{+1.5}_{-0.7}$\\
            \noalign{\smallskip}
            \hline
            \noalign{\smallskip}
                $\chi^2_\nu$ & $0.72 (105.5/146)$ \\[3pt] 
                $f_{0.3{\rm -}2}^{\rm obs}$
                         $~(10^{-13}~{\rm CGS})$
	                 & for A3: $1.5^{+0.2}_{-0.1}$\\[3pt]
	                 & for A4 $1.5^{+0.2}_{-0.2}$\\[3pt]
	                 & for A5: $1.2^{+0.1}_{-0.2}$\\[3pt]
                $f_{2{\rm -}10}^{\rm obs}$
                         $~(10^{-13}~{\rm CGS})$
	                 & for A3: $3.5^{+0.2}_{-0.2}$\\[3pt]
	                 & for A4 $3.5^{+0.2}_{-0.3}$\\[3pt]
	                 & for A5: $3.4^{+0.1}_{-0.2}$\\[3pt]
                $L^{\rm tot}_{0.3{\rm -}10}~(10^{39}~{\rm erg~s}^{-1})$ 
                        & $30.2^{+1.8}_{-1.5}$\\
            \noalign{\smallskip}
            \hline
         \end{tabular}
\end{center}
   \end{table}
\end{center}


\section{The mass of the accreting BH}

One of the unsolved mysteries regarding the nature of ULXs 
is the mass of the accreting BH. For NGC\,1365 X1, 
there is realistically no chance to obtain a dynamical 
mass: the optical counterpart is not detected in 
the OM or in any other archival image we could find, 
and at 19 Mpc (distance modulus $\approx 31.4$ mag), 
a companion star would in any case be too faint 
for optical spectroscopy. Therefore, we need to rely 
on X-ray data for the physical identification of the X-ray 
source. Two arguments can be used, based on the Eddington 
limit or on the X-ray spectral features. 

The 2006 outburst peaked at an X-ray flux corresponding 
to an isotropic luminosity 
$\approx 3 \times 10^{40}$ erg s$^{-1}$ in the 
$0.3$--$10$ keV band, identical to the isotropic luminosity 
measured by {\it ASCA} in 2005 January. The outburst 
sequence of our {\it Chandra} spectra suggests that 
the beginning of the decline is associated with an increase 
in ionized absorption (column density $\sim 10^{22}$ cm$^{-2}$, 
ionization parameter $\sim 100$), usually indicative 
of an outflow. We speculate that such outflow effectively 
blew away the X-ray emitting part of the inflow, causing 
the flux to drop by at least one order of magnitude. 
This behaviour is expected 
when the X-ray luminosity reaches or exceed the Eddington 
limit. This would suggest that the BH mass 
is $\la 200 M_{\odot}$. Coincidentally, the break in 
the ULX luminosity function is also found at 
an X-ray luminosity $\approx 3 \times 10^{40}$ erg s$^{-1}$ 
(Swartz et al.~2004; Gilfanov, Grimm \& Sunyaev 2004), 
perhaps suggesting that it corresponds to the Eddington limit 
of the most massive non-nuclear BHs in the local Universe.

It is also possible that the emission is not isotropic, 
owing to mild geometrical beaming, so that the true luminosity 
is reduced by a factor of a few (Fabrika \& Mescheryakov 2001; 
King et al.~2001). Furthermore, even for a standard 
Shakura-Sunyaev disk, the bolometric luminosity 
exceeds the classical Eddington luminosity by a factor 
$\approx 1 + (3/5) \ln (\dot{M}/\dot{M}_{\rm Edd})$ 
(Shakura \& Sunyaev 1973; Begelman, King \& Pringle 2006; 
Poutanen et al.~2007). In fact, a steady solution 
exists for accretion rates $\dot{M}/\dot{M}_{\rm Edd}$, 
in which the excess accreting matter is blown away 
in an outflow, which is optically thin along the disc axis 
but optically thick along the other directions. 
This provides a natural explanation for the scattering and 
collimation of the inner-disk emission along 
the disk axis (King \& Pounds 2003; Begelman et al.~2006; 
Poutanen et al.~2007). 
For a given observed flux, mild beaming and super-Eddington 
emission (Begelman 2006) reduce the required BH mass, 
making it consistent with the stellar-mass range.
 



The X-ray spectral analysis may provide an independent constraint 
to the BH mass and perhaps a better understanding 
of the source properties. If the source is in a thermal-dominant state 
(that is, dominated by a standard, optically-thick, geometrically-thin 
accretion disk), a tight correlation is predicted (and observed 
from Galactic BHs) between fitted inner-disk radius, 
bolometric disk luminosity and peak color temperature 
of the disk: $L_{\rm disk} \approx 4\pi R_{\rm in}^2 T_{\rm in}^4$. 
If we assume that $R_{\rm in} \approx R_{\rm isco} = \alpha GM/c^2$ 
(innermost stable circular orbit: $\alpha = 6$ 
in the Schwarzschild geometry, $\alpha = 1$ in 
the maximally-rotating Kerr geometry), 
the BH mass can be inferred from the previous relation, 
as $M \sim L_{\rm disk}^{1/2} T_{\rm in}^{-2}$
(Makishima et al.~2000; Ross, Fabian \& Ballantyne~2002).
In the case of NGC\,1365 X1, a candidate disk component 
is significantly detected at high signal-to-noise in the 
two long {\it XMM-Newton} observations E4 and E5 (Tables 5, 6). 
We extrapolate bolometric disk luminosities 
$\approx 3.3 \times 10^{39}$ erg s$^{-1}$ 
and $\approx 3.5 \times 10^{39}$ erg s$^{-1}$, for E4 and E5, 
and fitted inner-disk radii of $1555/\cos \theta$ km and 
$685/\cos \theta$ km respectively.
If we identify the apparent inner-disk radius 
with the innermost stable circular orbit, 
the BH mass would be $\ga 180 M_{\odot}$ (from E4) or 
$\ga 80 M_{\odot}$ (from E5).

This discrepancy between the two observations 
already hints that standard-disk relations may not 
be a reliable indicator for BH masses in ULXs.
More generally, we argue that such disk relations 
should not be directly applied to ULXs (see also 
Soria \& Kuncic 2007; Soria 2007). The reason is that the X-ray spectra 
of most ULXs (including NGC\,1365 X1) appear to be dominated by 
a Comptonization component (phenomenologically modelled 
by a power-law or broken power-law), not by thermal disk 
emission.
The presence of a dominant Comptonized component implies 
that most of the accretion power has been removed from the disk 
via non-radiative processes, and transferred to a jet or a corona. 
This necessarily affects the direct disk emission, in particular 
from the inner region, decreasing its luminosity and temperature. 
A similar phenomenon is observed from the Galactic BH 
XTE J1550$-$564 in its very high state (Kubota \& Done 2004; Done \& Kubota 2006).
A possible physical mechanism for a non-radiative, vertical extraction 
of power from the disk via magnetic torques was proposed 
by Kuncic \& Bicknell (2004). 
Its effect is to flatten 
the disk temperature profile at small radii; as a result, 
the peak of the disk emission occurs at larger 
radii ($\gg R_{\rm isco}$) 
and lower temperatures, leading to an over-estimate 
of the BH mass. Another possibility is that for very high 
accretion rates, the inner disk 
is completely replaced or covered by a moderately 
optically-thick corona, responsible for the power-law-like 
emission (Roberts 2007). In this case, too, the fitted inner-disk radius 
is not the innermost stable circular orbit: it is instead 
the transition radius between outer (standard) disk and 
modified inner region of the inflow, much larger 
than $R_{\rm isco}$.
Unfortunately, none of the brightest ULXs has ever been seen 
settling down into a disk-dominated state (unlike for example 
XTE J1550$-$564). So, we do not have an independent measurement 
of their innermost stable orbit when the standard disk 
is not affected by Comptonization.

In this scenario, we can use the fitted values of $R_{\rm in}$ and 
$T_{\rm in}$ to estimate the BH mass only 
if we can quantify how the transition radius $R_{\rm in}$ 
moves outwards from the innermost stable orbit, 
as a function of increasing $\dot{M}/\dot{M}_{\rm Edd}$ (Soria 2007). 
A possible suggestion is that such transition 
radius $\approx$ spherization radius (Shakura \& 
Sunyaev 1973), in which case 
$R_{\rm in} \sim (\dot{M}/\dot{M}_{\rm Edd}) R_{\rm isco}$ 
and $T_{\rm in} \sim (\dot{M}/\dot{M}_{\rm Edd})^{-1/2} T_\ast$, 
where $T_\ast$ would be the peak colour temperature 
of a standard disk extended to $R_{\rm isco}$. 
A more comprehensive discussion of the relation 
between transition radius and accretion rate is left 
to further work (Soria \& Kuncic~2007b, in preparation). 
Here we mention only one very simple argument: 
if the emitted luminosity in the disk 
component is $\sim 1/f$ times the total luminosity 
(with $f \sim 3$--$20$ for ``canonical'' ULXs, 
Stobbart et al.~2006), we speculate that the 
standard disk radiates most of the accretion power 
for $R \ga R_{\rm in} \approx fR_{\rm isco}$, while the accretion power 
released at $R_{\rm isco} \la R \la fR_{\rm isco}$ 
contributes to the Comptonized component.
This provides a lower limit for the transition radius, 
because the radiative efficiency of the Comptonizing medium 
is less than that of the disk, and part 
of the accretion power from the inner region may also be 
carried away as mechanical luminosity of an outflow.
For NGC\,1365 X1, we found that about $3/4$ of the X-ray luminosity 
was in the power-law component during the E4 and E5 
observations, which suggests that $> 3/4$ of the 
accretion power was extracted from the disk or inner inflow 
via non-radiative processes. 
(We also found that this fraction must have 
increased to $> 90\%$ during the 2006 outburst).
Taken at face value, this suggests that 
the fitted disk radius $R_{\rm in} \ga 4 R_{\rm isco}$. 
Thus, we argue that the X-ray spectral properties 
are consistent with BH masses $\sim 50$--$100 M_{\odot}$, 
even in the absence of beaming.

Can we reduce this mass estimate even further? 
We need to consider that even the most luminous disks detected 
in the thermal dominant or very high states 
of stellar-mass BHs 
are always at least a factor of three fainter than 
the (isotropic) disk luminosities inferred for 
NGC\,1365 X1 and other bright ULXs. 
No Galactic BH has been seen with a disk luminosity 
$\approx 3 \times 10^{39}$ erg s$^{-1}$ as measured in X1.
This suggests that the BH mass in this and other similar 
ULXs may still be a few times higher than in Galactic BHs, 
unless ULX luminosities are systemetically enhanced 
by mild beaming while those of the brightest 
Galactic BHs are not.

\section{Conclusions and speculations}

We have investigated the intriguing behaviour of 
a ULX in the spiral galaxy NGC\,1365 (in the 
Fornax cluster), characterized by high X-ray luminosity 
(peaking at $\approx 3 \times 10^{40}$ erg s$^{-1}$) 
and order-of-magnitude variability over timescales 
of weeks and months. We studied the system with 
{\it XMM-Newton} and {\it Chandra}, at different luminosities 
between 2002 and 2006. 

The X-ray spectrum of NGC\,1365 X1 is always dominated by 
a broad component well modelled as a power-law. 
The power-law photon index  
is $\Gamma \approx 1.6$--$1.9$, consistent with 
the 1995 {\it ASCA} observations. In fact, when 
an additional thermal component is also taken into 
account, the slope may be even higher, $\Gamma \approx 1$, 
at least in the $1$--$5$ keV range, with a possible break 
at higher energies. A soft excess consistent with 
disk-blackbody emission is significantly detected 
in the {\it XMM-Newton} spectra with higher signal-to-noise ratio. 
It contributes $\approx 1/4$ of the observed X-ray flux 
when the ULX has a $0.3$--$10$ keV luminosity 
$\approx 10^{40}$ erg s$^{-1}$, but it is not detected, 
and must contribute $< 10\%$ of the flux, when 
the ULX is near the outburst peak, 
at luminosities three times higher. 
This --- and similar findings for other bright ULXs --- suggests 
that one of the fundamental physical properties of this class 
of systems is that a large fraction of accretion power 
is extracted via non-thermal processes (for example, 
magnetic torques) rather than being directly radiated 
by an accretion disk, and that such fraction increases 
at higher accretion rates. A likely consequence is that 
the temperature profile of the inner accretion disk may be 
flatter than that of a standard disk; alternatively, the inner 
disk may be entirely replaced or covered by a Comptonizing region.
In all cases, the direct disk emission component would appear 
cooler and truncated at larger radii. This may lead 
to an overestimate of the central BH mass from simple 
spectral fitting (intermediate-mass BH scenario).
A similar phenomenon has been seen in at least 
one Galactic BH, XTE J1550$-$564 (Done \& Kubota 2006), 
and may be triggered when $\dot{M} \gg \dot{M}_{\rm Edd}$.

It is sometimes noted that the spectral state behaviour
of ULXs does not appear to follow the same pattern 
of canonical state transitions extensively 
studied in stellar-mass BHs (Section 1). 
We suggest that we can now start to outline a general 
sequence of spectral states applicable to both stellar-mass BHs 
and ULXs. Accreting BHs of any mass are 
in a low/hard state at low accretion rates 
($\dot{M} \la 0.01 \dot{M}_{\rm Edd}$), when the inner disk 
is truncated at large radii because not enough mass 
is flowing inwards. They are found in a soft (thermal-dominant) 
state at intermediate accretion rates 
($0.1 \dot{M}_{\rm Edd} \la \dot{M} \la 1 \dot{M}_{\rm Edd}$), 
when the disk reaches the innermost stable orbit 
and radiates all the accretion power. They might 
switch to a high/hard state at higher accretion rates 
($\dot{M} \ga \dot{M}_{\rm Edd}$) when most of the power 
begins to be extracted from the inner disk via 
non-radiative mechanisms (Poynting flux or mechanical power), 
and most of the emerging photons are coming from the Comptonizing 
medium. For $\dot{M} \gg \dot{M}_{\rm Edd}$ the Comptonizing corona 
may become optically thick or replace the disk altogether.
The open question is why ULXs seem to spend most of their active 
time in the high/hard state, rather than in the thermal 
dominant state. The reason may be related either to their higher 
BH mass or, more likely, to a consistently higher mass 
transfer rate from their massive donor stars.

An exceptional, serendipitous result of our {\it Chandra} sequence 
of observations in 2006 April is that we caught the rise, 
peak and decline of a strong but short-lived outburst 
($e$-folding timescale $\approx 3$ d). This flare may be similar 
to those observed in the power-law-dominated very high state 
of Galactic BHs. There is evidence of 
increased absorption from ionized gas (column 
density $\sim 10^{22}$ cm$^{-2}$, ionization 
parameter $\sim 100$) as the flux starts to decline. 
This might be due to the onset of an outflow as the source 
exceeded its Eddington limit. 
We speculate that the relatively low ionization parameter 
of the absorber is due to partial shielding of the wind from 
direct X-ray irradiation.

Finally, we discussed two possible arguments, based on the X-ray 
luminosity and spectral features, to constrain the mass 
of the accreting BH. We argue that the most likely range 
(consistent with both arguments)
is $\approx 50$--$150 M_{\odot}$ even without 
taking into account the possibility of mild geometrical 
beaming or super-Eddington emission (which would further 
reduce the required BH mass).

\section*{Acknowledgments}
We thank Martin Elvis, Zdenka Kuncic and Aneta Siemiginowska 
for valuable comments.  
This work was partially supported by NASA contract NAS 8-03060 
({\it Chandra} X-ray Center), {\it Chandra} grant G06-7102X 
(PI: GR) and {\it XMM-Newton} grant NAG 13161 (PI: GF). 
RS acknowledges support from an OIF Marie Curie 
Fellowship No.~509329

\end{document}